\documentstyle[emlines,titlepage]{article}

\begin{document}
\mbox{}\\
\mbox{}\\
\baselineskip 18pt
\large
\begin{center}
{\bf
Turbulent behaviour in magnetic hydrodynamics is not universal
    }
\end{center}
\mbox{}\\
\mbox{}\\

\begin{center}

{ D. Wolchenkov}
\end{center}

\begin{center}
{\it Department of Theoretical Physics, State University of St. Petersburg,

Ul'yanovskaya 1, Petrodvoretc, St. Petersburg, Russia}
\end{center}

\mbox{}\\
\mbox{}\\
\begin{center}
{\bf Abstract}
\end{center}

  A short distance
expansion  method (SDE) that is well known in the quantum field theory
for analysis of turbulent  behaviour  of stochastic magnetic \\
hydrodynamics of incompressible  conductive fluid is applied. As a result
is shown that  in an inertial range  the  turbulent spectra  of magnetic
 hydrodynamics depend  on a scale of arising of curls.

\mbox{}\\
\mbox{}\\

\section{ Introduction }

\ \ \ \ \ The methods of quantum  field theory   are successfully applied
for  description of the critical phenomena and developed turbulence recently.
This approach has  an important advantage before the classical one.
For example it enables one  to renormalize all the correlation
functions of a model as well as to define their various asymptotics.

A stochastic problem of magnetic hydrodynamics (MHD) of incompressible \\
conductive fluid with  external random force $ f^\varphi$ and
rotor of random current $f^\theta$
\begin{equation}
\label{1}
{\cal D}_t  v_i  =  \nu \Delta v_i - \partial_i p
+(\theta \partial) \theta_i + f^{v}_i,\quad {\cal D}_t =
\partial_t + v \partial,
\end{equation}
$$
{\cal D}_t  \theta_i  =  \nu' \Delta \theta_i +(\theta \partial)
v_i + f^{\theta}_i
$$
is equivalent to a theory of four fields with generating functional
 of renormalized correlation functions (Green functions)
$$
G(A_{\phi})=\int {\cal D}\Phi \det M \exp (S_{R}(\Phi)+\Phi A_{\phi}),
\quad\Phi=\{\varphi, \varphi',\theta ,\theta'\}
$$
in which the renormalized functional  of action is
\begin{equation}
\label{2}
S _R= \frac12 g_1 \nu^3 M^{2 \varepsilon} \varphi' D
^{\varphi\varphi} \varphi' +
 \frac12 g_2 \nu^3 M^{2a \varepsilon} \theta' D
^{\theta\theta} \theta' +
  g_3 \nu^3 M^{(1+a) \varepsilon} \varphi'  D
^{\varphi \theta} \theta' +
\end{equation}
$$+ \varphi' \bigl [ -\partial _t  \varphi + Z _1\nu \Delta \varphi - (
\varphi \partial) \varphi + Z _3 (\theta \partial)\theta  \bigr] +
 \theta' \bigl [ -\partial _t  \theta + Z _2\nu u\Delta \theta  - (
\varphi \partial) \theta  +  (\theta \partial)\varphi  \bigr] $$
(necessary integration on $ { \bf x } $ and $t$  as well as summation on
repeated badges are implied). Here the fields ${\bf \varphi}({\bf x}, t) $
 and $ { \bf \theta } ( { \bf x }, t) $  are both transversal (the vector
field of velocity $ {\bf \varphi } $  due to incompressibility of  fluid
($\partial _i\varphi _i =0$) and the pseudovector field $ { \bf\theta } $
as  connected with  transversal field of a magnetic induction $ { \bf B }:$
${ \bf \theta } = { \bf B } / \sqrt { 4\pi\varrho }$
($\varrho$  is a density of fluid,  $\nu$ is a viscosity, $p$ is a pressure).
 We shall use the dimensionless constant $\lambda$ (it means  inverted
number of Prandtl) instead of
 $\nu'=c^2 / 4\pi\sigma$ ( here $\sigma$ is a conductivity, $c$ is the
velocity of light) by this way $\nu'=\lambda\nu$.

Under construction of the model  existence of an inertial range is implied
therein  a real external
 energy  pumping (correlators of  the random forces) can be
simulated by  exponential model of $\delta$-function.
\begin{equation}
\label{3}
 D^{vv}_{is}  = {g_1}_0 \nu^3_0 P_{is}d_{vv} ,
 \quad
 D^{\theta\theta}_{is}  = {g_2}_0 \nu^3_0 P_{is}d_{\theta\theta} ,
 \quad
  D^{v\theta}_{is}  = {g_3}_0 \nu^3_0 \varepsilon_{ism}  k_m
d_{v\theta} ,
\end{equation}
$$d_{vv}= k^{4-d-2\varepsilon},
\quad d_{\theta\theta} =k^{4-d-2a\varepsilon},\quad
d_{v\theta}= k^{3-d-(1+ a)\varepsilon}.$$

In momentum-frequency representation (from frequency the correlators
$D_{is}$ do not depend) $P_{is}$  is the transversal projector,
 $d$ means the  dimension of  space (completely antisymmetric pseudotensor
  $\varepsilon _ { ism } $ is determined only at $d=3$).

The factors $g_0$ in correlators are played a role of charges;
 the positive constant $a$ is an arbitrary parameter.
The parameter $\varepsilon$ serves for construction of \\ decomposition
of correlation functions, as the physical value of $\varepsilon$
the  $\varepsilon_p = 2$  is considered that simulates energy pumping
from the large-scale movements of  fluid.

 The model (\ref{1} -\ref{3}) was investigated in \cite {1} in the
first order on $\varepsilon$ with the help of recursive renormalization
group method. It is shown that in the system two different regimes of
 developed turbulence can be realized: "kinetic" and
"magnetic" that are connected with existence of two infrared (IR) stable
 fixed  points of RG transformation. In \cite{1}  the critical
asymptotics of spectral density of energy $<{\bf\varphi}({\bf k})
 {\bf\varphi}({\bf -k})>$ and $<{\bf\theta}({\bf k})
{\bf\theta}({\bf -k})>$ were determined. In \cite {2} more
 general formulation of the problem with  inclusion of the cross correlator
 of random force $D^ { \varphi\theta } $ was considered by the  quantum field
RG method. The renormalization constants $Z_{i}$ were calculated there in
the first order of $ g_{\alpha} $:
$$
Z _1 =
1-\frac{g_1d(d-1)}{4B\varepsilon}-
\frac{g_2(d^2+d-4)}{4Ba\lambda^2\varepsilon},
\quad
Z _3 =
1+\frac {g_1}{B\lambda\varepsilon}-\frac {g_2}{Ba\lambda^2\varepsilon},
$$
$$
Z _2 =
1-\frac{g_1(d
+2)(d-1)}{2B\lambda(\lambda+1)\varepsilon}-
\frac{g_2(d+2)(d-3)}{2Ba\lambda^2(\lambda+1)\varepsilon};
$$
here $B=d(d+2){(4 \pi)}^{d/2}\Gamma (d/2)$ and  $\Gamma (x)$ is
the gamma-function.

Renormalizability of the  model   was   proven,  and situation
in the charging space  of IR-stable fixed points corresponding to
 kinetic and magnetic regimes were found:
\begin{equation}
\label{4}
{g'_1}_{*}\equiv\frac{{g _1}_{*}}{B\lambda_{*}} =\frac{\varepsilon (1+\lambda_{*})}{15},\quad
{g _2}_{*} =0,\quad \lambda_{*}=  \frac{\sqrt{43/3}-1}{2},
\end{equation}
( the region of stability of the point is  $a < 1.16.$ )
\begin{equation}
\label{5}
{g _1}_{*} =\lambda_{*}=0   ,\quad
{g'_2}_{*}\equiv\frac{{g _2}_{*}}{B{\lambda_{*}}^2} =a\varepsilon
\end{equation}
( this one is stable at $a\geq 0.25$).
In paper \cite{3}   features  of  a scaling behaviour in the model  were
 investigated, and  critical dimensions of fields and parameters of the
theory are found out in the  both critical regimes.

We shall be interested in dependence
of characteristics of developed turbulence (the correlation functions)
in the inertial range from conditions of arising of large-scale curls.
In the theory of developed turbulence it is supposed, that the energy
pumping into the inertial range is executed by the vortices of a large size
$\Lambda$. We shall take into account  this scale having supplied the
model (\ref{1} -\ref{3} ) by an infrared mass parameter
 $m\equiv 1 / \Lambda$ to consider
the relative corrections to  developed turbulent spectra with the help
of SDE method.

According to this method following the operator decomposition is fair:
 \begin{equation}
\label{6}
{ \bf \phi } ( { \bf x _1 }, t ){ \bf \phi } ( { \bf x _2 }, t) \simeq
\sum _ {i } c _i ( { \bf r } ) F _i ( { \bf x }, t);
\end{equation}
where $x\equiv ( x _1 + x _2 )/ 2$, $r\equiv  x_1 -x _2$,
$F _i$  are various local averages (composite operators).
The averaging of (\ref{6}) yields the asymptotics for   pair correlation
 functions  at $mr\to 0$
\begin{equation}
\label{7}
< { \bf \phi } ( { \bf x _1 }, t ){ \bf \phi } ( { \bf x _2 }, t)
>\simeq\sum _
{ i } c _i ( { \bf r } ) a _i m^ { \Delta _ { F _i }};
\end{equation}
here $\Delta _ { F _i } $ are the critical dimensions of the composite
operators; $a_i$  means some constants. On decomposition (\ref {7}) we
conclude,  that from the point of view of an opportunity of transition to
a massless theory ( $m\to 0$ ) in the inertial range the operators with
a  negative critical dimension  are dangerous.

\section{Critical dimensions of the composite operators in the model of
magnetic hydrodynamics}

 The renormalized  operators are defined by the formula
 $F_i=Z_ { ik } F^R_k ( \Phi_R ).$ On the known matrix $Z_ { ik } $ a
  matrix of anomalous dimensions $\gamma_ { ik } = ( Z^ { -1 } ) _ { ij } {
\cal D } _MZ_ { jk }$ is calculated and then  the matrix of critical
dimensions $\Delta_ { ik } = ( d^k_F ) _ { ik } + \Delta_\omega ( d^\omega_F )
_ { ik } + \gamma_ { ik }$ where $\Delta _\omega$  designates critical
dimension of frequency, and $d^k_F$,  $d^\omega_F$ are the momentum
 and frequency  canonical dimensions of $F$; they are being determined from
  requirement of  frequency and momentum dimensionless of terms  of the
action functional.

The particular critical dimensions are  eigenvalues  of the  matrix
 $\Delta_ { ik }$. They correspond to linear combinations of composite
 operators $L_i (F^R ) =U_ { ik } F^R_k$ which diagonalize
the matrix $\Delta_ { ik }$.

We shall consider  the dimensions of  elementary composite operators of
the MHD model: $\phi_i\phi_k$, $\phi'\phi$, as well as
vector operators $ ( \partial \phi\phi ) _i$ with various transpositions
of badges.

The tensor $\phi_i\phi_k$ is a sum of two independent tensors
\begin{equation}
\label{8}
\frac 1d \phi^2 \delta_{ik},
\quad\quad\quad \phi_i\phi_k-\frac 1d\phi^2\delta_{ik}.
\end{equation}
Convolution of the first operator on badges yields family of scalar
operators $\phi\phi$
$$
F _1=\frac 12 v _iv  _i;\quad
F _2=\frac 12 \theta _i\theta  _i;\quad
F _3= \theta _iv  _i.
$$
Trace of the second expression in (\ref{8}) is equal to  zero.

 Essential property of the theory (\ref{1} - \ref{3}) is Galileian
   invariancy; it was used in \cite {4} for investigation of the composite
 operators in a problem of usual stochastic hydrodynamics.
A non-stationary Galileian transformations of the fields
$$ \varphi_a ({\bf x},t)= \varphi  ({\bf x +u}(t),t)-{\bf a}(t);\quad
\varphi' _a ({\bf x},t)= \varphi'  ({\bf x +u}(t),t);$$
$$
\theta _a ({\bf x},t)= \theta  ({\bf x +u}(t),t);  \quad
\theta' _a ({\bf x},t)= \theta' ({\bf x +u}(t),t);
$$
 (a parameter of transformation - $ { \bf v } (t) $  is
the vector function dependent only  from a time   well decreasing at
 $ | t | \to \infty$  and $ { \bf u } (t) = \int _ { -\infty } ^ { t } dt
 { \bf v } ( t' ) $ )
realized in  the Ward identity  leads \cite{5} to
\begin{equation}
\label{9}
\int dx\{a _{0\alpha}[\frac {\partial F _\alpha}{\partial \varphi _s}-{
\frac {\partial F _\alpha}{\partial (\partial _t \varphi _k)}}  \partial _s
\varphi _k] + \partial _t[a _{0\alpha} \frac {\partial F _\alpha}{\partial
(\partial _s \varphi _k)}]\} <\infty.
\end{equation}
Here $a _ { 0\alpha } $  are the nonrenormalized functions of
 sources of family of  operators $F _\alpha$, and symbol
 ''$ < \infty$ '' means that the considered functional is finite.

 We shall assign $G_1= v_iv_k / 2$ and $G_2=\theta_i\theta_k / 2$.
The operator $G_3=v_i\theta_k$ is pseudotensor so it doesn't mix
 with the  first two operators due to renormalization.
 $G_1$ is noninvariant of the Galileian transformations  so  the formula
 (\ref{9}) permits to prove  that $G _1 $   is finite ($Z_ { 11 } =1 $),
and that it doesn't mix  to the operator $G _2$ in
 proceeding of renormalization, hence $\quad Z_ { 21 } =0$
(one can  say, that the operator $G _1$  aren't being renormalized and
doesn't mix to  the Galileian invariant  operator $G _2$ since  $G _1$ is not
 Galileian invariant).
 Besides, from (\ref{9}) the absence of divergences in  diagrams with
the composite operator $G_3$ (it is noninvariant of Galileian
transformations also) is  followed.
 Then from the definition of renormalized composite operators it is easy
to get the expression:  $Z_ { 33 } =Z_ { v_i\theta_k }
 =Z_\theta^ { -1 }.$

 Thus, the matrix of renormalization constants  ${\bf Z}_{\alpha\delta}$
 of the operators $G_i$ is
\begin{displaymath}
{\bf Z} _{\alpha\beta}=
\left(\begin{array}{ccc}
1&Z _{12}&0\\
0&Z _{22}&0\\
0&0&Z _\theta ^{-1}
\end{array}\right).
\end{displaymath}
Unknown elements of the matrix one can calculate with the help of
standard diagrams with the following propagators
\begin{eqnarray}
\label{10}
<\varphi _i({\bf k},t)\varphi' _j({\bf-k},0)> _0=
P _{ij}e^{-\nu k^2 t}\theta  (t); \, \nonumber\\
<\varphi _i ({\bf k},t)\varphi _j({\bf-k},0)> _0=
\frac12 P _{ij}g \nu^2k^{2-d-2\varepsilon} M^{2\varepsilon}e^{-\nu k^2 |t|};
\, \nonumber\\
<\theta _i({\bf k},t)\theta' _j({\bf-k},0)> _0=
P _{ij}\theta (t)e ^{-\nu \lambda k^2 t}; \, \nonumber\\
<\theta _i({\bf k},t)\theta _j({\bf-k},0)> _0=
\frac12  P_{ij}\frac{g'\nu^2k^{2-d-2a\varepsilon}}{\lambda}
M^{2a\varepsilon}e^{-\nu\lambda k^2|t|};
\end{eqnarray}
$$<\theta _i({\bf k},t)\varphi _j({\bf-k},0)> _0=
\varepsilon _{isj}\frac{g'' \nu^2k^{1-d-(1+a)\varepsilon}}{(1+\lambda)}
M^{(1+a)
\varepsilon}k _s
[e^{-\lambda\nu k^2t}\theta (t)+e^{\nu k^2 t}\theta (-t)];$$
Account of the renormalization constants for the scalar operator we
execute  having curtailed the badges $\delta _ { ik }.$

The calculations
in one-loop  approximation under diagrams specified on a fig. 1 of the
appropriate constants  $Z _ { \alpha\delta } $ in the theory (\ref{2})
yelds a renormalized action  for the generating functional of correlation
functions  with the composite operators $F_\alpha$: $\bar S _R= S _R +
a_{0\alpha}Z_{\alpha\delta}F_\delta^R (Z_\phi\phi)$; here the
 renormalization constants are
\begin{eqnarray}
\label{11}
{Z^{\varphi\theta}} _{12} =
-\frac{\lambda(d
+2)(d-1)}{2(\lambda+1)}(\frac {g'_1}{\varepsilon}-\frac{g' _2}{a\varepsilon});
\quad {Z^{\theta^2}} _{22}Z _\theta ^2 = 1+
\frac {(d-1)(d+2)}{2(\lambda+1)}(\frac
 {g'_1}{\varepsilon}-\frac{g' _2}{a\varepsilon});\,
\nonumber \\
{Z^{\varphi_{i}\theta_{j}}}_ {12} =
-\frac \lambda{2(\lambda+1)}(\frac {g'_1}{\varepsilon}-
\frac{g' _2}{a\varepsilon}),
\quad
{Z^{\theta_{i}\theta_{j}}}_{22} Z _\theta ^2 = 1+\frac 1{2(\lambda+1)}(\frac
 {g'_1}{\varepsilon}-\frac{g' _2}{a\varepsilon}).
\end{eqnarray}

As far as in the fixed points  of RG transformation  some values of the
 charges approach to zero  in avoidance of trivialization of  asymptotics
 in all orders of the $\varepsilon$-decomposition it is useful to
redefine the fields as follows:
\begin{equation}
\label{12}
\theta \rightarrow\sqrt{g_2\nu ^3}M ^{a\varepsilon}\theta;
\quad\quad\quad
v\rightarrow\sqrt{g_1\nu^3} M^\varepsilon v;
\end{equation}
$$
\theta' \rightarrow\theta'/\sqrt{g_2\nu ^3}M ^{a\varepsilon};
\quad\quad\quad\varphi'\rightarrow\varphi'/\sqrt{g_1\nu^3} M^\varepsilon.
$$
The transformations  don't change positions of the fixed points
and  values of the renormalization constants (the canonical dimensions of
the  operators  $F _\alpha$ vary only).

 The critical dimensions of the operators $G_i$ calculated on the constants
 $Z _ { \alpha\alpha } $ in the first order of  $\varepsilon$ and at
 any $d$  are listed in table I.

It should be noticed that the value of  $\Delta _1$ is exact.
 The constant $Z_{12}$ defines an admixture to  $G _1$ of the operator
$G _2$.
Considering  (\ref{2}) in a condition of the kinetic regime with the
fields of (\ref{12})  we have
$G_2\to g_2 G_2$ and taking into account that in  the kinetic fixed point
(\ref{4}) $g _2^ { * } =0$ it's easy to show  that just the operators
 $G _1,\quad G _2$ (instead of their mixture) have the dimension
$\Delta _i$ in this point.

The constant $Z_{12}\sim O(\lambda)$ which is responsible for mixing of the
operators disappears in the magnetic point.

 We shall consider elements of a matrix $Z_{ik}$ correspond to
renormalization of the set of scalar and vector composite operators
$\phi'\phi$ and $ ( \partial \phi\phi ) _i$.

At this set there are the operators reducing to  a total differential
 $\partial ( \phi\phi )$ with various transpositions of badges.
Renormalization of them  is equivalent to renormalization
of the operators ${\bf\phi}{\bf\phi}$  have considered above.
The critical dimensions is being appropriated to these operators surpass
dimensions located in table I per unit of.

The operators of a $\phi'\phi$-type don't mix to any other operators
 and aren't  being renormalized because of 1-irreducible diagrams
that is responsible for mixing of these operators with the other are
equal to zero as far as they contain  cycles of advancing lines.
A structure of interactions in the  (\ref{2}) provides removal
 from each diagram a one derivative on each external line of
 a $\phi'$-type that effectively lowers an index of divergence of
the diagrams. Therefore, in the minimal subtraction (MS) scheme of
renormalization that is appropriate to these operators the diagonal
elements are $Z_ { \alpha\alpha } =1$,  and all nondiagonal ones are equal
to  zero. Thus, $\Delta_{\phi'\phi}=3.$

The remaining vector operators $F_1=v_iv^2$ and $F_2=v\theta ^2$
are  true tensors, and $F_3=\theta_i\theta^2$,
 $\quad F_4=\theta v^2$  are  pseudotensors; these pairs of
 the operators are being renormalized independently from each other.
The operators $F_{1-2}$ are noninvariant  to Galileian
transformations, thereof, it is easy to approve  \cite{5} the finiteness of
the  operator $v^3$ as well as that it doesn't mix to $v{\theta^2}$ due to
renormalization. Thus, $Z_ { 11 } =1,\quad Z_ { 21 } =0.$

 Similarly, the Galileian invariant operator $F_3$ can't mix to the
 operator $F_4$ that  means $Z_{34}=0.$

For definition of remaining elements of the matrix  $Z_ { ik } $ in
a one-loop \\ approximation it is necessary to consider the diagrams that is
shown  on a fig. 2. It gives
 $$Z_{12}=-3 \frac{\lambda (d+2)(d-1)}
 {(\lambda+1)}(\frac{g'_1}{\varepsilon}-\frac{g'_2}{a\varepsilon})
 ,\quad Z_{33}Z^3_\theta=1+ 3 \frac{(d+2)(d-1)}
 {(\lambda+1)}(\frac{g'_1}{\varepsilon}-\frac{g'_2}{a\varepsilon}), $$
 \begin{equation}
\label{13}
 Z_{22}Z^2_\theta= 1+\left( \frac{(d-1)(d+2)}{(\lambda+1)}-1\right)
 (\frac{g'_1}{\varepsilon}-\frac{g'_2}{a\varepsilon}), \label{16}
 \end{equation}
 $$\quad Z_{43}Z_\theta=- \frac{\lambda (d+2)(d-1)}
 {(\lambda+1)}(\frac{g'_1}{\varepsilon}-\frac{g'_2}{a\varepsilon}),\quad
 Z_{44}Z_\theta=1- \frac{g'_1}{\varepsilon}+\frac{g'_2}{a\varepsilon},$$
 The matrix of renormalization  constants  of this family of the operators
  has a block triangular form, so the critical dimensions
  are determined  by the diagonal elements $Z_{\alpha\alpha}.$

The values of critical dimensions calculated on (\ref{13}) are listed
in table 2.

The nondiagonal elements  $Z_ { 12 } $ and $Z_ { 43 } $ define
an admixture of the operators $F_1$ and $F_4$ to $F_2$ and $F_3.$
Taking into account that in the kinetic mode $F_1= { g_1 } ^ { 3 / 2 }
v^3$, $F_2=g_1^ { 1 / 2 } g_2 v\theta^2$ $F_3= { g_2 } ^ { 3 / 2 } \theta ^3$,
and $F_4= g_1 { g_2 } ^ { 1 / 2 } \theta v^2$ and that  $ {g_2 } _ { * } =0$
in the fixed point (\ref{5}) it is possible to assert
the certain critical dimensions are belonged to the  operators,
 instead of their linear combinations. In the magnetic point (\ref{6})
 $Z_ { 12 } =Z_ { 34 } =0.$

\section{Discussion of the  results }

It is important to note that any of the operators considered  can't
participate as  amendments for phenomenological equations of MHD.
 The operators are being possessed of the essential critical dimensions
 don't contain  auxiliary fields $\phi'$, but the operators of
a $\phi'\phi$-type (a function of response) are inessential
 and  don't satisfy the requirements  of Galileian invariancy also.

The operators of canonical dimension $d=3$ define new nonanalytic corrections
  to the spectra of developed turbulence  that was found in \cite{3}.
According to SDE method such corrections for a pair correlation function
 $ < \phi_1\phi_2>$ can be represented as follows:
 \begin{equation}
 <\phi_1({\bf k},t)\phi_2({\bf -k},t)>=
 Ak^{-d -\Delta _{\phi_1}-\Delta _{\phi_2}}\left(1+\sum_i b_i \left(\frac
 {m}{k}\right)^{ \Delta_{F_i}}\right);
 \label{14}
  \end{equation}
(here $\Delta_\phi$ are the critical dimensions of the fields).

As far as in the inertial range the following estimation is correct
 $ m / k \ll 1$, the formula (\ref{14}) results to nonanalyticities
 in a case of $ \Delta_ { F_i } < 0.$ At the real value of $\varepsilon =2$
the critical  dimensions of  $G_2 \quad ( a>1 / 2 $, $F_1,\quad F_2
 \quad ( a>0.243 ), \quad F_3,\quad F_4 \quad ( a>0.82 ) $ in the
kinetic point become negative.
 The dimension of  $F _4$ at $a>3 / 4$ is negative  in the magnetic
 point also.

  Thanking to Galileian invariancy of the theory  we can
refer  to the results of  \cite{4} where the terms  of the sum
of (\ref{14})  for static correlation functions
 connected with the Galileian noninvariant  composite operators
 were proven to  yield not a contribution. Those in our case
are $F_1,\quad F_2$,
 and $F_4.$ They can participate in the decomposition (\ref{16}) only for
 dynamic correlation functions. The Galileian invariant operator $F_3$
 gives the contribution  in the sum in all the cases.

Comparing of the dimensions of the composite operators of families
 $\phi\phi$ and $( \partial \phi\phi ) _i$ at the real value
 $\varepsilon=2$ one can see that the set
 $ ( \partial \phi\phi ) _i$ appears  more essential in the scaling
range (so in  the kinetic regime $\Delta_ { \theta^2 } >
\Delta_ { v\theta^2 },\Delta_ { \theta^3 },\quad
\Delta_ { v^2 } >\Delta_ { v^3 } $ ).
One can assume that the tendency of growth  of essentiality of  operators
is being demonstrated by the elementary operators in the model
 of magnetic  hydrodynamics  will be saved  for  more complex  operators.

 The results received by us testify that in the inertial range in MHD
 the correlation functions depend on  the external integral turbulent
 scale $ m.$  Thus,  the behaviour in the  model is not universal.

 This important result can be checked experimentally besides the particular
 values  of critical dimensions of the composite operators calculated
here can be measured on an experiment too that will be served certainly
 to check of the  offered theory.

\section{ Acknowledgments}

The author is grateful to M. Yu. Nalimov and L. Ts. Adzhemyan for useful
discussion at preparation of the paper as well as to mayorate of
 St.-Petersburg (Russia ) for rendered support.

 \newpage
\vspace{15cm}

\ \ \
{\bf Table I}
\ \ \

\begin{tabular}{||c|c|c||}
                                    \hline
\multicolumn{3}{|c|}{The critical dimensions of
 the composite operators of  $\phi_{i}\phi_j$ type}\\ \hline
\multicolumn{1}{|c|}{ Operator}&
\multicolumn{1}{|c|}{Kinetic point}&
\multicolumn{1}{|c|}{Magnetic point}\\
                                \hline  \hline
\it
$(G_1)_{ij}$&$2-4/3\varepsilon$&$2$\\ \hline
\it
$(G_2)_{ij}$&$2-2(a-3/10)\varepsilon$&$2+3a\varepsilon$ \\ \hline
\it
$(G_3)_{ij}$ & $2-(a+1/3)\varepsilon$ & $2+a\varepsilon$ \\  \hline
\it
$(G_1)_{ij}\delta_{ij} $ & $2-4/3\varepsilon$& $2$ \\ \hline
\it
$(G_2)_{ij}\delta_{ij} $ & $2-2a\varepsilon$& $2+12a\varepsilon$ \\ \hline
\it
$(G_3)_{ij}\delta_{ij} $ & $2-(a+1/3)\varepsilon$& $2+a\varepsilon$ \\ \hline
\hline
\end{tabular}

\ \ \

\ \ \

 \newpage
\vspace{15cm}

\ \ \
{\bf Table II}
\ \ \

\begin{tabular}{||c|c|c||}
                                    \hline
\multicolumn{3}{|c|}{The critical dimensions of
 the composite operators of  $(\partial\phi\phi)_i$ type}\\ \hline
\multicolumn{1}{|c|}{ Operator}&
\multicolumn{1}{|c|}{Kinetic point}&
\multicolumn{1}{|c|}{Magnetic point}\\
                                \hline  \hline
\it
$F_1$&$3-2\varepsilon$&$3$\\ \hline
\it
$F_2$&$3-2(a+0.507)\varepsilon$&$3+18a\varepsilon$ \\ \hline
\it
$F_3$ & $3-3(a+1)\varepsilon$ & $3+60a\varepsilon$ \\  \hline
\it
$F_4$ & $3-(a+0.68)\varepsilon$& $3-2a\varepsilon$ \\ \hline
\hline
\end{tabular}

\ \ \

\ \ \

\newpage
\vspace{15cm}
\unitlength=1.00mm
\special{em:linewidth 0.4pt}
\linethickness{0.4pt}
\begin{picture}(154.00,37.00)
\put(154.00,15.00){\makebox(0,0)[cc]{$,$}}
\put(19.00,19.00){\rule{2.00\unitlength}{1.00\unitlength}}
\emline{21.00}{20.00}{1}{38.00}{37.00}{2}
\emline{21.00}{19.00}{3}{38.00}{2.00}{4}
\emline{30.00}{10.00}{5}{30.00}{29.00}{6}
\put(55.00,19.00){\rule{2.00\unitlength}{1.00\unitlength}}
\emline{57.00}{20.00}{7}{74.00}{37.00}{8}
\emline{57.00}{19.00}{9}{74.00}{2.00}{10}
\emline{66.00}{10.00}{11}{66.00}{29.00}{12}
\put(82.00,19.00){\rule{2.00\unitlength}{1.00\unitlength}}
\emline{84.00}{20.00}{13}{101.00}{37.00}{14}
\emline{84.00}{19.00}{15}{101.00}{2.00}{16}
\emline{93.00}{10.00}{17}{93.00}{29.00}{18}
\put(116.00,19.00){\rule{2.00\unitlength}{1.00\unitlength}}
\emline{118.00}{20.00}{19}{135.00}{37.00}{20}
\emline{118.00}{19.00}{21}{135.00}{2.00}{22}
\emline{127.00}{10.00}{23}{127.00}{29.00}{24}
\emline{24.00}{25.00}{25}{27.00}{26.00}{26}
\emline{27.00}{26.00}{27}{27.00}{24.00}{28}
\emline{27.00}{15.00}{29}{29.00}{11.00}{30}
\emline{29.00}{11.00}{31}{26.00}{11.00}{32}
\emline{62.00}{12.00}{33}{65.00}{11.00}{34}
\emline{65.00}{11.00}{35}{64.00}{14.00}{36}
\emline{65.00}{22.00}{37}{66.00}{26.00}{38}
\emline{66.00}{26.00}{39}{67.00}{22.00}{40}
\emline{86.00}{24.00}{41}{90.00}{26.00}{42}
\emline{90.00}{26.00}{43}{88.00}{22.00}{44}
\emline{89.00}{16.00}{45}{91.00}{12.00}{46}
\emline{91.00}{12.00}{47}{87.00}{14.00}{48}
\emline{121.00}{14.00}{49}{124.00}{13.00}{50}
\emline{124.00}{13.00}{51}{123.00}{16.00}{52}
\emline{126.00}{20.00}{53}{127.00}{23.00}{54}
\emline{127.00}{23.00}{55}{128.00}{19.00}{56}
\put(20.00,24.00){\makebox(0,0)[cc]{$\theta$}}
\put(25.00,29.00){\makebox(0,0)[cc]{$\theta'$}}
\put(32.00,35.00){\makebox(0,0)[cc]{$\theta$}}
\put(33.00,26.00){\makebox(0,0)[cc]{$\varphi$}}
\put(33.00,14.00){\makebox(0,0)[cc]{$\varphi$}}
\put(20.00,14.00){\makebox(0,0)[cc]{$\theta$}}
\put(25.00,9.00){\makebox(0,0)[cc]{$\theta'$}}
\put(32.00,3.00){\makebox(0,0)[cc]{$\theta$}}
\put(44.00,19.00){\makebox(0,0)[cc]{$,$}}
\put(56.00,24.00){\makebox(0,0)[cc]{$\theta$}}
\put(62.00,29.00){\makebox(0,0)[cc]{$\theta$}}
\put(68.00,35.00){\makebox(0,0)[cc]{$\theta$}}
\put(69.00,26.00){\makebox(0,0)[cc]{$\varphi'$}}
\put(69.00,14.00){\makebox(0,0)[cc]{$\varphi$}}
\put(56.00,14.00){\makebox(0,0)[cc]{$\theta$}}
\put(62.00,9.00){\makebox(0,0)[cc]{$\theta'$}}
\put(68.00,3.00){\makebox(0,0)[cc]{$\theta$}}
\put(83.00,24.00){\makebox(0,0)[cc]{$\varphi$}}
\put(89.00,29.00){\makebox(0,0)[cc]{$\varphi'$}}
\put(94.00,35.00){\makebox(0,0)[cc]{$\theta$}}
\put(96.00,26.00){\makebox(0,0)[cc]{$\theta$}}
\put(96.00,14.00){\makebox(0,0)[cc]{$\theta$}}
\put(83.00,14.00){\makebox(0,0)[cc]{$\varphi$}}
\put(89.00,9.00){\makebox(0,0)[cc]{$\varphi'$}}
\put(94.00,3.00){\makebox(0,0)[cc]{$\theta$}}
\put(75.00,19.00){\makebox(0,0)[cc]{$,$}}
\put(105.00,19.00){\makebox(0,0)[cc]{$,$}}
\put(118.00,24.00){\makebox(0,0)[cc]{$\varphi$}}
\put(124.00,29.00){\makebox(0,0)[cc]{$\varphi$}}
\put(129.00,35.00){\makebox(0,0)[cc]{$\theta$}}
\put(130.00,27.00){\makebox(0,0)[cc]{$\theta'$}}
\put(130.00,14.00){\makebox(0,0)[cc]{$\theta$}}
\put(118.00,14.00){\makebox(0,0)[cc]{$\varphi$}}
\put(124.00,9.00){\makebox(0,0)[cc]{$\varphi'$}}
\put(129.00,3.00){\makebox(0,0)[cc]{$\theta$}}
\end{picture}

\begin{center}
{\bf fig. 1} The set of one-loop 1-irreducible diagrams are
responsible for the nontrival renormalization constances of the
 composite operators of a $ \phi_{i}\phi_{j}$ type.
\end{center}

\newpage
\vspace{3cm}
\unitlength=0.50mm
\special{em:linewidth 0.4pt}
\linethickness{0.4pt}
\begin{picture}(288.00,298.00)
\put(22.00,279.00){\makebox(0,0)[cc]{$Z _{12}$:}}
\emline{44.00}{279.00}{1}{55.00}{279.00}{2}
\emline{57.00}{278.00}{3}{74.00}{261.00}{4}
\put(55.00,278.00){\rule{2.00\unitlength}{2.00\unitlength}}
\emline{57.00}{280.00}{5}{75.00}{298.00}{6}
\emline{69.00}{292.00}{7}{69.00}{266.00}{8}
\emline{68.00}{276.00}{9}{69.00}{271.00}{10}
\emline{69.00}{271.00}{11}{70.00}{276.00}{12}
\emline{64.00}{289.00}{13}{68.00}{291.00}{14}
\emline{68.00}{291.00}{15}{66.00}{287.00}{16}
\emline{97.00}{279.00}{17}{110.00}{279.00}{18}
\put(110.00,278.00){\rule{2.00\unitlength}{2.00\unitlength}}
\emline{112.00}{280.00}{19}{129.00}{297.00}{20}
\emline{112.00}{278.00}{21}{130.00}{260.00}{22}
\emline{125.00}{265.00}{23}{125.00}{293.00}{24}
\emline{120.00}{290.00}{25}{124.00}{292.00}{26}
\emline{123.00}{291.00}{27}{122.00}{288.00}{28}
\emline{122.00}{270.00}{29}{124.00}{266.00}{30}
\emline{124.00}{266.00}{31}{121.00}{268.00}{32}
\put(172.00,279.00){\makebox(0,0)[cc]{$Z _{33}$:}}
\emline{194.00}{279.00}{33}{205.00}{279.00}{34}
\emline{207.00}{278.00}{35}{224.00}{261.00}{36}
\put(205.00,278.00){\rule{2.00\unitlength}{2.00\unitlength}}
\emline{207.00}{280.00}{37}{225.00}{298.00}{38}
\emline{219.00}{292.00}{39}{219.00}{266.00}{40}
\emline{218.00}{276.00}{41}{219.00}{271.00}{42}
\emline{219.00}{271.00}{43}{220.00}{276.00}{44}
\emline{214.00}{289.00}{45}{218.00}{291.00}{46}
\emline{218.00}{291.00}{47}{216.00}{287.00}{48}
\emline{247.00}{279.00}{49}{260.00}{279.00}{50}
\put(260.00,278.00){\rule{2.00\unitlength}{2.00\unitlength}}
\emline{262.00}{280.00}{51}{279.00}{297.00}{52}
\emline{262.00}{278.00}{53}{280.00}{260.00}{54}
\emline{275.00}{265.00}{55}{275.00}{293.00}{56}
\emline{270.00}{290.00}{57}{274.00}{292.00}{58}
\emline{273.00}{291.00}{59}{272.00}{288.00}{60}
\emline{272.00}{270.00}{61}{274.00}{266.00}{62}
\emline{274.00}{266.00}{63}{271.00}{268.00}{64}
\put(102.00,230.00){\makebox(0,0)[cc]{$Z _{43}$:}}
\emline{124.00}{230.00}{65}{135.00}{230.00}{66}
\emline{137.00}{229.00}{67}{154.00}{212.00}{68}
\put(135.00,229.00){\rule{2.00\unitlength}{2.00\unitlength}}
\emline{137.00}{231.00}{69}{155.00}{249.00}{70}
\emline{149.00}{243.00}{71}{149.00}{217.00}{72}
\emline{148.00}{227.00}{73}{149.00}{222.00}{74}
\emline{149.00}{222.00}{75}{150.00}{227.00}{76}
\emline{144.00}{240.00}{77}{148.00}{242.00}{78}
\emline{148.00}{242.00}{79}{146.00}{238.00}{80}
\emline{177.00}{230.00}{81}{190.00}{230.00}{82}
\put(190.00,229.00){\rule{2.00\unitlength}{2.00\unitlength}}
\emline{192.00}{231.00}{83}{209.00}{248.00}{84}
\emline{192.00}{229.00}{85}{210.00}{211.00}{86}
\emline{205.00}{216.00}{87}{205.00}{244.00}{88}
\emline{200.00}{241.00}{89}{204.00}{243.00}{90}
\emline{203.00}{242.00}{91}{202.00}{239.00}{92}
\emline{202.00}{221.00}{93}{204.00}{217.00}{94}
\emline{204.00}{217.00}{95}{201.00}{219.00}{96}
\put(44.00,272.00){\makebox(0,0)[cc]{$v$}}
\put(56.00,284.00){\makebox(0,0)[cc]{$v$}}
\put(63.00,292.00){\makebox(0,0)[cc]{$v'$}}
\put(78.00,293.00){\makebox(0,0)[cc]{$\theta$}}
\put(74.00,287.00){\makebox(0,0)[cc]{$\theta$}}
\put(76.00,273.00){\makebox(0,0)[cc]{$\theta'$}}
\put(74.00,265.00){\makebox(0,0)[cc]{$\theta$}}
\put(62.00,264.00){\makebox(0,0)[cc]{$v$}}
\put(54.00,271.00){\makebox(0,0)[cc]{$v$}}
\put(97.00,272.00){\makebox(0,0)[cc]{$v$}}
\put(108.00,271.00){\makebox(0,0)[cc]{$v$}}
\put(117.00,265.00){\makebox(0,0)[cc]{$v'$}}
\put(130.00,265.00){\makebox(0,0)[cc]{$\theta$}}
\put(129.00,273.00){\makebox(0,0)[cc]{$\theta$}}
\put(129.00,286.00){\makebox(0,0)[cc]{$\theta$}}
\put(133.00,294.00){\makebox(0,0)[cc]{$\theta$}}
\put(112.00,285.00){\makebox(0,0)[cc]{$v$}}
\put(120.00,294.00){\makebox(0,0)[cc]{$v'$}}
\put(196.00,273.00){\makebox(0,0)[cc]{$\theta$}}
\put(206.00,285.00){\makebox(0,0)[cc]{$\theta$}}
\put(207.00,272.00){\makebox(0,0)[cc]{$\theta$}}
\put(212.00,265.00){\makebox(0,0)[cc]{$\theta$}}
\put(213.00,292.00){\makebox(0,0)[cc]{$\theta'$}}
\put(229.00,294.00){\makebox(0,0)[cc]{$\theta$}}
\put(224.00,285.00){\makebox(0,0)[cc]{$v$}}
\put(225.00,273.00){\makebox(0,0)[cc]{$v'$}}
\put(224.00,265.00){\makebox(0,0)[cc]{$\theta$}}
\put(251.00,272.00){\makebox(0,0)[cc]{$\theta$}}
\put(262.00,284.00){\makebox(0,0)[cc]{$\theta$}}
\put(260.00,272.00){\makebox(0,0)[cc]{$\theta$}}
\put(268.00,292.00){\makebox(0,0)[cc]{$\theta'$}}
\put(267.00,266.00){\makebox(0,0)[cc]{$\theta'$}}
\put(281.00,272.00){\makebox(0,0)[cc]{$v$}}
\put(281.00,285.00){\makebox(0,0)[cc]{$v$}}
\put(283.00,295.00){\makebox(0,0)[cc]{$\theta$}}
\put(280.00,264.00){\makebox(0,0)[cc]{$\theta$}}
\put(211.00,237.00){\makebox(0,0)[cc]{$\theta$}}
\put(210.00,223.00){\makebox(0,0)[cc]{$\theta$}}
\put(214.00,246.00){\makebox(0,0)[cc]{$\theta$}}
\put(210.00,215.00){\makebox(0,0)[cc]{$\theta$}}
\put(179.00,221.00){\makebox(0,0)[cc]{$\theta$}}
\put(191.00,221.00){\makebox(0,0)[cc]{$v$}}
\put(192.00,236.00){\makebox(0,0)[cc]{$v$}}
\put(201.00,245.00){\makebox(0,0)[cc]{$v'$}}
\put(200.00,216.00){\makebox(0,0)[cc]{$v'$}}
\put(156.00,216.00){\makebox(0,0)[cc]{$\theta$}}
\put(157.00,244.00){\makebox(0,0)[cc]{$\theta$}}
\put(156.00,237.00){\makebox(0,0)[cc]{$\theta$}}
\put(154.00,225.00){\makebox(0,0)[cc]{$\theta'$}}
\put(125.00,223.00){\makebox(0,0)[cc]{$\theta$}}
\put(135.00,222.00){\makebox(0,0)[cc]{$v$}}
\put(137.00,236.00){\makebox(0,0)[cc]{$v$}}
\put(144.00,243.00){\makebox(0,0)[cc]{$v'$}}
\put(140.00,216.00){\makebox(0,0)[cc]{$v$}}
\put(10.00,182.00){\makebox(0,0)[cc]{$Z _{22}$:}}
\emline{32.00}{182.00}{97}{43.00}{182.00}{98}
\emline{45.00}{181.00}{99}{62.00}{164.00}{100}
\put(43.00,181.00){\rule{2.00\unitlength}{2.00\unitlength}}
\emline{45.00}{183.00}{101}{63.00}{201.00}{102}
\emline{57.00}{195.00}{103}{57.00}{169.00}{104}
\emline{56.00}{179.00}{105}{57.00}{174.00}{106}
\emline{57.00}{174.00}{107}{58.00}{179.00}{108}
\emline{52.00}{192.00}{109}{56.00}{194.00}{110}
\emline{56.00}{194.00}{111}{54.00}{190.00}{112}
\emline{85.00}{182.00}{113}{98.00}{182.00}{114}
\put(98.00,181.00){\rule{2.00\unitlength}{2.00\unitlength}}
\emline{100.00}{183.00}{115}{117.00}{200.00}{116}
\emline{100.00}{181.00}{117}{118.00}{163.00}{118}
\emline{113.00}{168.00}{119}{113.00}{196.00}{120}
\emline{108.00}{193.00}{121}{112.00}{195.00}{122}
\emline{111.00}{194.00}{123}{110.00}{191.00}{124}
\emline{110.00}{173.00}{125}{112.00}{169.00}{126}
\emline{112.00}{169.00}{127}{109.00}{171.00}{128}
\put(116.00,190.00){\makebox(0,0)[cc]{$\theta$}}
\put(116.00,173.00){\makebox(0,0)[cc]{$\theta$}}
\put(118.00,197.00){\makebox(0,0)[cc]{$v$}}
\put(118.00,167.00){\makebox(0,0)[cc]{$\theta$}}
\put(87.00,178.00){\makebox(0,0)[cc]{$\theta$}}
\put(100.00,177.00){\makebox(0,0)[cc]{$v$}}
\put(100.00,188.00){\makebox(0,0)[cc]{$\theta$}}
\put(109.00,197.00){\makebox(0,0)[cc]{$\theta'$}}
\put(106.00,167.00){\makebox(0,0)[cc]{$v'$}}
\put(64.00,168.00){\makebox(0,0)[cc]{$\theta$}}
\put(64.00,197.00){\makebox(0,0)[cc]{$v$}}
\put(60.00,190.00){\makebox(0,0)[cc]{$v$}}
\put(61.00,177.00){\makebox(0,0)[cc]{$v'$}}
\put(31.00,174.00){\makebox(0,0)[cc]{$\theta$}}
\put(45.00,176.00){\makebox(0,0)[cc]{$\theta$}}
\put(45.00,188.00){\makebox(0,0)[cc]{$v$}}
\put(52.00,195.00){\makebox(0,0)[cc]{$v'$}}
\put(50.00,168.00){\makebox(0,0)[cc]{$\theta$}}
\emline{155.00}{182.00}{129}{168.00}{182.00}{130}
\put(168.00,181.00){\rule{2.00\unitlength}{2.00\unitlength}}
\emline{170.00}{183.00}{131}{187.00}{200.00}{132}
\emline{170.00}{181.00}{133}{188.00}{163.00}{134}
\emline{183.00}{168.00}{135}{183.00}{196.00}{136}
\emline{178.00}{193.00}{137}{182.00}{195.00}{138}
\emline{181.00}{194.00}{139}{180.00}{191.00}{140}
\emline{180.00}{173.00}{141}{182.00}{169.00}{142}
\emline{182.00}{169.00}{143}{179.00}{171.00}{144}
\put(186.00,190.00){\makebox(0,0)[cc]{$v$}}
\put(186.00,173.00){\makebox(0,0)[cc]{$v$}}
\put(188.00,197.00){\makebox(0,0)[cc]{$\theta$}}
\put(188.00,167.00){\makebox(0,0)[cc]{$\theta$}}
\put(157.00,178.00){\makebox(0,0)[cc]{$v$}}
\put(169.00,175.00){\makebox(0,0)[cc]{$\theta$}}
\put(170.00,188.00){\makebox(0,0)[cc]{$\theta$}}
\put(179.00,197.00){\makebox(0,0)[cc]{$\theta'$}}
\put(176.00,166.00){\makebox(0,0)[cc]{$\theta'$}}
\emline{224.00}{182.00}{145}{237.00}{182.00}{146}
\put(237.00,181.00){\rule{2.00\unitlength}{2.00\unitlength}}
\emline{239.00}{183.00}{147}{256.00}{200.00}{148}
\emline{239.00}{181.00}{149}{257.00}{163.00}{150}
\emline{252.00}{168.00}{151}{252.00}{196.00}{152}
\emline{247.00}{193.00}{153}{251.00}{195.00}{154}
\emline{250.00}{194.00}{155}{249.00}{191.00}{156}
\emline{249.00}{173.00}{157}{251.00}{169.00}{158}
\emline{251.00}{168.00}{159}{248.00}{170.00}{160}
\put(255.00,190.00){\makebox(0,0)[cc]{$v$}}
\put(256.00,175.00){\makebox(0,0)[cc]{$v$}}
\put(257.00,197.00){\makebox(0,0)[cc]{$v$}}
\put(257.00,167.00){\makebox(0,0)[cc]{$\theta$}}
\put(226.00,178.00){\makebox(0,0)[rc]{$\theta$}}
\put(236.00,175.00){\makebox(0,0)[cc]{$\theta$}}
\put(239.00,188.00){\makebox(0,0)[cc]{$v$}}
\put(248.00,197.00){\makebox(0,0)[cc]{$v'$}}
\put(243.00,165.00){\makebox(0,0)[cc]{$\theta'$}}
\emline{32.00}{128.00}{161}{43.00}{128.00}{162}
\emline{45.00}{127.00}{163}{62.00}{110.00}{164}
\put(43.00,127.00){\rule{2.00\unitlength}{2.00\unitlength}}
\emline{45.00}{129.00}{165}{63.00}{147.00}{166}
\emline{57.00}{141.00}{167}{57.00}{115.00}{168}
\emline{56.00}{125.00}{169}{57.00}{120.00}{170}
\emline{57.00}{120.00}{171}{58.00}{125.00}{172}
\emline{52.00}{138.00}{173}{56.00}{140.00}{174}
\emline{56.00}{140.00}{175}{54.00}{136.00}{176}
\put(64.00,114.00){\makebox(0,0)[rc]{$\theta$}}
\put(66.00,142.00){\makebox(0,0)[cc]{$v$}}
\put(61.00,134.00){\makebox(0,0)[cc]{$\theta$}}
\put(62.00,122.00){\makebox(0,0)[cc]{$\theta'$}}
\put(33.00,122.00){\makebox(0,0)[cc]{$\theta$}}
\put(45.00,122.00){\makebox(0,0)[lc]{$v$}}
\put(45.00,134.00){\makebox(0,0)[cc]{$\theta$}}
\put(52.00,141.00){\makebox(0,0)[cc]{$\theta'$}}
\put(49.00,114.00){\makebox(0,0)[cb]{$v$}}
\emline{85.00}{128.00}{177}{96.00}{128.00}{178}
\emline{98.00}{127.00}{179}{115.00}{110.00}{180}
\put(96.00,127.00){\rule{2.00\unitlength}{2.00\unitlength}}
\emline{98.00}{129.00}{181}{116.00}{147.00}{182}
\emline{110.00}{141.00}{183}{110.00}{115.00}{184}
\emline{109.00}{125.00}{185}{110.00}{120.00}{186}
\emline{110.00}{120.00}{187}{111.00}{125.00}{188}
\emline{105.00}{138.00}{189}{109.00}{140.00}{190}
\emline{109.00}{140.00}{191}{107.00}{136.00}{192}
\put(117.00,114.00){\makebox(0,0)[cc]{$v$}}
\put(120.00,143.00){\makebox(0,0)[cc]{$\theta$}}
\put(113.00,136.00){\makebox(0,0)[cc]{$\theta$}}
\put(115.00,121.00){\makebox(0,0)[cc]{$\theta'$}}
\put(87.00,121.00){\makebox(0,0)[cc]{$\theta$}}
\put(98.00,122.00){\makebox(0,0)[cc]{$\theta$}}
\put(98.00,134.00){\makebox(0,0)[cc]{$v$}}
\put(105.00,141.00){\makebox(0,0)[cc]{$v'$}}
\put(103.00,113.00){\makebox(0,0)[cc]{$\theta$}}
\emline{155.00}{128.00}{193}{166.00}{128.00}{194}
\emline{168.00}{127.00}{195}{185.00}{110.00}{196}
\put(166.00,127.00){\rule{2.00\unitlength}{2.00\unitlength}}
\emline{168.00}{129.00}{197}{186.00}{147.00}{198}
\emline{180.00}{141.00}{199}{180.00}{115.00}{200}
\emline{179.00}{125.00}{201}{180.00}{120.00}{202}
\emline{180.00}{120.00}{203}{181.00}{125.00}{204}
\emline{175.00}{138.00}{205}{179.00}{140.00}{206}
\emline{179.00}{140.00}{207}{177.00}{136.00}{208}
\put(187.00,114.00){\makebox(0,0)[cc]{$v$}}
\put(190.00,143.00){\makebox(0,0)[cc]{$\theta$}}
\put(183.00,136.00){\makebox(0,0)[cc]{$v$}}
\put(185.00,121.00){\makebox(0,0)[cc]{$v'$}}
\put(157.00,121.00){\makebox(0,0)[cc]{$\theta$}}
\put(168.00,122.00){\makebox(0,0)[cc]{$v$}}
\put(168.00,134.00){\makebox(0,0)[cc]{$\theta$}}
\put(175.00,141.00){\makebox(0,0)[cc]{$\theta'$}}
\put(173.00,114.00){\makebox(0,0)[cc]{$v$}}
\emline{222.00}{128.00}{209}{233.00}{128.00}{210}
\emline{235.00}{127.00}{211}{252.00}{110.00}{212}
\put(233.00,127.00){\rule{2.00\unitlength}{2.00\unitlength}}
\emline{235.00}{129.00}{213}{253.00}{147.00}{214}
\emline{247.00}{141.00}{215}{247.00}{115.00}{216}
\emline{246.00}{125.00}{217}{247.00}{120.00}{218}
\emline{247.00}{120.00}{219}{248.00}{125.00}{220}
\emline{242.00}{138.00}{221}{246.00}{140.00}{222}
\emline{246.00}{140.00}{223}{244.00}{136.00}{224}
\put(254.00,114.00){\makebox(0,0)[cc]{$\theta$}}
\put(257.00,143.00){\makebox(0,0)[cc]{$\theta$}}
\put(253.00,136.00){\makebox(0,0)[cc]{$v$}}
\put(252.00,123.00){\makebox(0,0)[cc]{$v'$}}
\put(223.00,122.00){\makebox(0,0)[cc]{$v$}}
\put(235.00,122.00){\makebox(0,0)[cc]{$\theta$}}
\put(235.00,134.00){\makebox(0,0)[cc]{$\theta$}}
\put(242.00,141.00){\makebox(0,0)[cc]{$\theta'$}}
\put(240.00,114.00){\makebox(0,0)[cc]{$\theta$}}
\put(10.00,77.00){\makebox(0,0)[cc]{$Z _{44}$:}}
\emline{32.00}{77.00}{225}{43.00}{77.00}{226}
\emline{45.00}{76.00}{227}{62.00}{59.00}{228}
\put(43.00,76.00){\rule{2.00\unitlength}{2.00\unitlength}}
\emline{45.00}{78.00}{229}{63.00}{96.00}{230}
\emline{57.00}{90.00}{231}{57.00}{64.00}{232}
\emline{56.00}{74.00}{233}{57.00}{69.00}{234}
\emline{57.00}{69.00}{235}{58.00}{74.00}{236}
\emline{52.00}{87.00}{237}{56.00}{89.00}{238}
\emline{56.00}{89.00}{239}{54.00}{85.00}{240}
\emline{85.00}{77.00}{241}{98.00}{77.00}{242}
\put(98.00,76.00){\rule{2.00\unitlength}{2.00\unitlength}}
\emline{100.00}{78.00}{243}{117.00}{95.00}{244}
\emline{100.00}{76.00}{245}{118.00}{58.00}{246}
\emline{113.00}{63.00}{247}{113.00}{91.00}{248}
\emline{108.00}{88.00}{249}{112.00}{90.00}{250}
\emline{111.00}{89.00}{251}{110.00}{86.00}{252}
\emline{110.00}{68.00}{253}{112.00}{64.00}{254}
\emline{113.00}{62.00}{255}{110.00}{64.00}{256}
\put(118.00,85.00){\makebox(0,0)[cc]{$v$}}
\put(118.00,69.00){\makebox(0,0)[cc]{$v$}}
\put(122.00,93.00){\makebox(0,0)[cc]{$v$}}
\put(118.00,62.00){\makebox(0,0)[cc]{$v$}}
\put(87.00,71.00){\makebox(0,0)[cc]{$\theta$}}
\put(98.00,70.00){\makebox(0,0)[cc]{$v$}}
\put(100.00,83.00){\makebox(0,0)[cc]{$v$}}
\put(109.00,92.00){\makebox(0,0)[cc]{$v'$}}
\put(107.00,60.00){\makebox(0,0)[cc]{$v'$}}
\put(64.00,63.00){\makebox(0,0)[cc]{$v$}}
\put(67.00,93.00){\makebox(0,0)[cc]{$v$}}
\put(62.00,85.00){\makebox(0,0)[rb]{$v$}}
\put(62.00,72.00){\makebox(0,0)[cc]{$v'$}}
\put(33.00,70.00){\makebox(0,0)[cc]{$\theta$}}
\put(43.00,70.00){\makebox(0,0)[cc]{$v$}}
\put(45.00,83.00){\makebox(0,0)[cc]{$v$}}
\put(52.00,90.00){\makebox(0,0)[cc]{$v'$}}
\put(50.00,63.00){\makebox(0,0)[cc]{$v$}}
\emline{155.00}{77.00}{257}{168.00}{77.00}{258}
\put(168.00,76.00){\rule{2.00\unitlength}{2.00\unitlength}}
\emline{170.00}{78.00}{259}{187.00}{95.00}{260}
\emline{170.00}{76.00}{261}{188.00}{58.00}{262}
\emline{183.00}{63.00}{263}{183.00}{91.00}{264}
\emline{178.00}{88.00}{265}{182.00}{90.00}{266}
\emline{181.00}{89.00}{267}{180.00}{86.00}{268}
\emline{180.00}{68.00}{269}{182.00}{64.00}{270}
\emline{182.00}{63.00}{271}{179.00}{65.00}{272}
\put(189.00,85.00){\makebox(0,0)[cc]{$v$}}
\put(189.00,71.00){\makebox(0,0)[cc]{$v$}}
\put(192.00,93.00){\makebox(0,0)[cc]{$v$}}
\put(188.00,62.00){\makebox(0,0)[cc]{$\theta$}}
\put(157.00,71.00){\makebox(0,0)[cc]{$v$}}
\put(169.00,70.00){\makebox(0,0)[cc]{$\theta$}}
\put(170.00,83.00){\makebox(0,0)[cc]{$v$}}
\put(179.00,92.00){\makebox(0,0)[cc]{$v'$}}
\put(176.00,61.00){\makebox(0,0)[cc]{$\theta'$}}
\emline{224.00}{77.00}{273}{237.00}{77.00}{274}
\put(237.00,76.00){\rule{2.00\unitlength}{2.00\unitlength}}
\emline{239.00}{78.00}{275}{256.00}{95.00}{276}
\emline{239.00}{76.00}{277}{257.00}{58.00}{278}
\emline{252.00}{63.00}{279}{252.00}{91.00}{280}
\emline{247.00}{88.00}{281}{251.00}{90.00}{282}
\emline{250.00}{89.00}{283}{249.00}{86.00}{284}
\emline{249.00}{68.00}{285}{251.00}{64.00}{286}
\emline{251.00}{64.00}{287}{248.00}{66.00}{288}
\put(258.00,85.00){\makebox(0,0)[cc]{$\theta$}}
\put(258.00,71.00){\makebox(0,0)[cc]{$\theta$}}
\put(259.00,94.00){\makebox(0,0)[cc]{$v$}}
\put(257.00,62.00){\makebox(0,0)[cc]{$\theta$}}
\put(227.00,69.00){\makebox(0,0)[cc]{$v$}}
\put(238.00,70.00){\makebox(0,0)[cc]{$\theta$}}
\put(239.00,83.00){\makebox(0,0)[cc]{$v$}}
\put(248.00,92.00){\makebox(0,0)[cc]{$v'$}}
\put(245.00,63.00){\makebox(0,0)[cc]{$\theta'$}}
\emline{32.00}{23.00}{289}{43.00}{23.00}{290}
\emline{45.00}{22.00}{291}{62.00}{5.00}{292}
\put(43.00,22.00){\rule{2.00\unitlength}{2.00\unitlength}}
\emline{45.00}{24.00}{293}{63.00}{42.00}{294}
\emline{57.00}{36.00}{295}{57.00}{10.00}{296}
\emline{56.00}{20.00}{297}{57.00}{15.00}{298}
\emline{57.00}{15.00}{299}{58.00}{20.00}{300}
\emline{52.00}{33.00}{301}{56.00}{35.00}{302}
\emline{56.00}{35.00}{303}{54.00}{31.00}{304}
\put(64.00,9.00){\makebox(0,0)[cc]{$v$}}
\put(70.00,39.00){\makebox(0,0)[cc]{$\theta$}}
\put(62.00,31.00){\makebox(0,0)[cc]{$\theta$}}
\put(62.00,18.00){\makebox(0,0)[cc]{$\theta'$}}
\put(34.00,18.00){\makebox(0,0)[cc]{$v$}}
\put(43.00,15.00){\makebox(0,0)[cc]{$\theta$}}
\put(45.00,29.00){\makebox(0,0)[cc]{$v$}}
\put(52.00,36.00){\makebox(0,0)[cc]{$v'$}}
\put(48.00,9.00){\makebox(0,0)[cc]{$\theta$}}
\emline{85.00}{23.00}{305}{96.00}{23.00}{306}
\emline{98.00}{22.00}{307}{115.00}{5.00}{308}
\put(96.00,22.00){\rule{2.00\unitlength}{2.00\unitlength}}
\emline{98.00}{24.00}{309}{116.00}{42.00}{310}
\emline{110.00}{36.00}{311}{110.00}{10.00}{312}
\emline{109.00}{20.00}{313}{110.00}{15.00}{314}
\emline{110.00}{15.00}{315}{111.00}{20.00}{316}
\emline{105.00}{33.00}{317}{109.00}{35.00}{318}
\emline{109.00}{35.00}{319}{107.00}{31.00}{320}
\put(117.00,9.00){\makebox(0,0)[cc]{$v$}}
\put(119.00,38.00){\makebox(0,0)[cc]{$\theta$}}
\put(116.00,31.00){\makebox(0,0)[cc]{$v$}}
\put(115.00,18.00){\makebox(0,0)[cc]{$v'$}}
\put(87.00,18.00){\makebox(0,0)[cc]{$v$}}
\put(96.00,15.00){\makebox(0,0)[cc]{$v$}}
\put(98.00,29.00){\makebox(0,0)[cc]{$\theta$}}
\put(104.00,38.00){\makebox(0,0)[cc]{$\theta'$}}
\put(103.00,9.00){\makebox(0,0)[cc]{$v$}}
\emline{155.00}{23.00}{321}{166.00}{23.00}{322}
\emline{168.00}{22.00}{323}{185.00}{5.00}{324}
\put(166.00,22.00){\rule{2.00\unitlength}{2.00\unitlength}}
\emline{168.00}{24.00}{325}{186.00}{42.00}{326}
\emline{180.00}{36.00}{327}{180.00}{10.00}{328}
\emline{179.00}{20.00}{329}{180.00}{15.00}{330}
\emline{180.00}{15.00}{331}{181.00}{20.00}{332}
\emline{175.00}{33.00}{333}{179.00}{35.00}{334}
\emline{179.00}{35.00}{335}{177.00}{31.00}{336}
\put(187.00,9.00){\makebox(0,0)[ct]{$\theta$}}
\put(190.00,39.00){\makebox(0,0)[cc]{$v$}}
\put(185.00,31.00){\makebox(0,0)[lb]{$\theta$}}
\put(185.00,18.00){\makebox(0,0)[cc]{$\theta'$}}
\put(156.00,17.00){\makebox(0,0)[cc]{$v$}}
\put(165.00,14.00){\makebox(0,0)[cc]{$v$}}
\put(168.00,29.00){\makebox(0,0)[cc]{$\theta$}}
\put(172.00,37.00){\makebox(0,0)[cc]{$\theta'$}}
\put(173.00,8.00){\makebox(0,0)[cc]{$v$}}
\emline{222.00}{23.00}{337}{233.00}{23.00}{338}
\emline{235.00}{22.00}{339}{252.00}{5.00}{340}
\put(233.00,22.00){\rule{2.00\unitlength}{2.00\unitlength}}
\emline{235.00}{24.00}{341}{253.00}{42.00}{342}
\emline{247.00}{36.00}{343}{247.00}{10.00}{344}
\emline{246.00}{20.00}{345}{247.00}{15.00}{346}
\emline{247.00}{15.00}{347}{248.00}{20.00}{348}
\emline{242.00}{33.00}{349}{246.00}{35.00}{350}
\emline{246.00}{35.00}{351}{244.00}{31.00}{352}
\put(254.00,9.00){\makebox(0,0)[rc]{$\theta$}}
\put(256.00,38.00){\makebox(0,0)[cc]{$v$}}
\put(251.00,31.00){\makebox(0,0)[cc]{$v$}}
\put(252.00,18.00){\makebox(0,0)[cc]{$v'$}}
\put(222.00,17.00){\makebox(0,0)[cc]{$v$}}
\put(233.00,15.00){\makebox(0,0)[cc]{$\theta$}}
\put(235.00,29.00){\makebox(0,0)[cc]{$v$}}
\put(242.00,38.00){\makebox(0,0)[cc]{$v'$}}
\put(240.00,8.00){\makebox(0,0)[cc]{$\theta$}}
\put(144.00,278.00){\makebox(0,0)[cc]{;}}
\put(288.00,278.00){\makebox(0,0)[cc]{;}}
\put(223.00,228.00){\makebox(0,0)[cc]{;}}
\put(270.00,125.00){\makebox(0,0)[cc]{;}}
\put(270.00,19.00){\makebox(0,0)[cc]{;}}
\end{picture}
\mbox{}\\
\mbox{}\\

\begin{center}
{\bf fig. 2} The set of one-loop 1-irreducible diagrams are
responsible for the nontrival renormalization constances of the
 composite operators of a $ (\partial\phi\phi)_{i}$ type.
\end{center}


\begin{thebibliography}{99}
\bibitem{1}
Fournier J.-P., Frish U. {\em Phys.Rev.}, 1983. Vol. A28,$N^0$ 2. P.1000.
\bibitem{2}
   L. Ts. Adzhemyan,   A.N. Vasil'ev,  M. Gnatich,  {\em Theor. and math.
physics},  1985, Vol.  64, $N^0$ 2,  p. 196.
\bibitem{3}
L. Ts. Adzhemyan,  D.Yu. Wolchenkov,  M. Yu. Nalimov,
{\em Theor. and math. physics}, Vol. 107, $N^0 1$, pp. 142-154.
\bibitem{4}
L. Ts. Adzhemyan,   N.V. Antonov, A.N. Vasil'ev,  {\em Journal of
experimental and theoretical physics},  1989,  Vol. { 95},
$N^{0} 4$, pp. 1272, (on Russian).
\bibitem{5}
   L. Ts. Adzhemyan,   A.N. Vasil'ev,  M. Gnatich,  {\em Theor. and math.
physics},  1988, Vol. { 74}, N2,  p. 180.
  \end{thebibliography}
 \end{document}